\documentclass[conference]{IEEEtran}
\IEEEoverridecommandlockouts
\usepackage[english]{babel}
\usepackage{amsthm}
\usepackage{titlesec}
\usepackage[ruled,vlined]{algorithm2e}
\usepackage{algcompatible}
\usepackage{graphicx}
\usepackage{float}
\usepackage{caption}
\usepackage{tikz}
\usepackage{changepage}
\usepackage[utf8]{inputenc}
\usepackage{pgfplots} 
\usepackage{pgfgantt}
\usepackage{pdflscape}
 \usepackage{relsize}
\usepackage[export]{adjustbox}
\pgfplotsset{compat=newest} 
\pgfplotsset{plot coordinates/math parser=false}
\pgfplotsset{compat=1.18}
\usepackage[justification=centering]{caption}
\usepackage{pgfplots}
\usetikzlibrary{spy}

\newtheorem{innercustomgeneric}{\customgenericname}
\providecommand{\customgenericname}{}
\newcommand{\newcustomtheorem}[2]{%
  \newenvironment{#1}[1]
  {%
   \renewcommand\customgenericname{#2}%
   \renewcommand\theinnercustomgeneric{##1}%
   \innercustomgeneric
  }
  {\endinnercustomgeneric}
}

\newcustomtheorem{customthm}{Theorem}
\newcustomtheorem{customlemma}{Lemma}
\newcustomtheorem{customprop}{Proposition}

\usepackage{lipsum}
\usepackage{amsmath}
\usepackage{acronym}
\usepackage{amssymb}
\usepackage{mathtools}
\usepackage{url}
\usepackage{graphicx}  
\usepackage{float}  
\newcommand{\yy}{\mathbf{y}}
\newcommand{\xx}{\mathbf{x}}
\newcommand{\uu}{\mathbf{u}}
\newcommand{\bb}{\mathbf{b}}
\newcommand{\zz}{\mathbf{z}}
\newcommand{\hh}{\mathbf{h}}

\newcommand{\pp}{\mathbf{p}}
\newcommand{\qq}{\mathbf{q}}
\newcommand{\rr}{\mathbf{r}}

\newcommand{\blkdiag}{\textrm{blkdiag}}

\newcommand{\Nrf}{{N_{\rm{RF}}}}

\newcommand{\ww}{\mathbf{w}}

\newcommand{\BB}{\mathbf{B}}
\newcommand{\UU}{\mathbf{U}}

\newcommand{\GG}{\mathbf{G}}
\newcommand{\YY}{\mathbf{Y}}
\newcommand{\NN}{\mathbf{N}}
\newcommand{\HH}{\mathbf{H}}
\newcommand{\ff}{\mathbf{f}}

\newcommand{\WW}{\mathbf{W}}

\newcommand{\FF}{\mathbf{F}}
\newcommand{\PP}{\mathbf{P}}
\newcommand{\XX}{\mathbf{X}}
\newcommand{\RR}{\mathbf{R}}
\newcommand{\DD}{\mathbf{D}}

\newcommand{\SSb}{\mathbf{S}}
\newcommand{\zzeta}{\boldsymbol{\zeta}}

\newcommand{\Tr}{\text{Tr}}
\newcommand{\diag}{\text{diag}}

\newcommand{\hermit}{\mathsf{H}}

\usepackage{accents}
\newcommand*{\dt}[1]{%
	\accentset{\mbox{\large .}}{#1}}
\newcommand*{\ddt}[1]{%
	\accentset{\mbox{\large ..}}{#1}}

\newcommand{\JJ}{\mathbf{J}}
\newcommand{\TT}{\mathbf{T}}

\newcommand{\AAb}{\mathbf{A}}

\newcommand{\atxdt}{\dt{\mathbf{a}}_{{\rmtx}} }
\newcommand{\atxdttilde}{\widetilde{\dt{\mathbf{a}}}_{{\rmtx}} }
\newcommand{\atxddttilde}{\widetilde{\ddt{\mathbf{a}}}_{{\rmtx}} }
\newcommand{\atxddt}{\ddt{\mathbf{a}}_{{\rmtx}} }

\newcommand{\norm}[1]{\left\lVert#1\right\rVert}

\newcommand{\rmtx}{{\rm{Tx}}}

\newcommand{\dd}{\mathbf{d}}

\newcommand{\ee}{\mathbf{e}}
\newcommand{\TTheta}{\boldsymbol{\Theta}}
\newcommand{\ttheta}{\boldsymbol{\theta}}

\newcommand{\vecc}{\text{vec}}

\newcommand{\eeta}{\boldsymbol{\eta}}

\newcommand{\aris}{\mathbf{a}_{\textrm{RIS}}}
\newcommand{\atx}{\mathbf{a}_\rmtx}
\newcommand{\rmt}{{\textrm{t}}}

\newcommand{\Mtx}{ M_{\textrm{Tx}}}
\newcommand{\Mtxz}{ M_{\textrm{Tx},z}}
\newcommand{\Mtxy}{ M_{\textrm{Tx},y}}
\newcommand{\Mrx}{M_{\textrm{Rx}}}
\newcommand{\tx}{{\textrm{Tx}}}

\acrodef{SISO}[SISO]{single-input single-output}
\acrodef{AP}[AP]{access point}
\acrodef{UE}[UE]{user equipment}
\acrodef{ULA}[ULA]{uniform linear array}
\acrodef{CPU}[CPU]{central processing unit}
\acrodef{FPP}[FPP]{Feasible-point pursuit}
\acrodef{LoS}[LoS]{line-of-sight}
\acrodef{NLoS}[NLoS]{non-line-of-sight}
\acrodef{RCS}[RCS]{radar cross section}
\acrodef{AoD}[AoD]{angle of departure}
\acrodef{AoA}[AoA]{angle of arrival}
\acrodef{CRB}[CRB]{Cramer-Rao bound}
\acrodef{FIM}[FIM]{Fisher information matrix}
\acrodef{AN}[AN]{artificial noise}
\acrodef{SINR}[SINR]{signal-to-interference-plus-noise ratio}
\acrodef{SNR}[SNR]{signal-to-noise ratio}
\acrodef{QoS}[QoS]{quality of service}
\acrodef{SDR}[SDR]{semi-definite relaxation}
\acrodef{SDP}[SDP]{semi-definite relaxation}
\acrodef{ISAC}[ISAC]{integrated sensing and communications}
\acrodef{PLS}[PLS]{physical layer security}
\acrodef{SIC}[SIC]{successive interference cancellation}
\acrodef{CSI}[CSI]{channel state information}
\acrodef{MUI}[MUI]{multi-user interference}
\acrodef{RIS}[RIS]{Reconfigurable intelligent surface}
\acrodef{AO}[AO]{alternating optimization}
\acrodef{SIMO}[SIMO]{Single Input Multiple Output}
\acrodef{MISO}[MISO]{multiple-intput single output}
\acrodef{MIMO}[MIMO]{multiple-input multiple-output}
\acrodef{MU}{multi-user}
\acrodef{BS}[BS]{base station}
\acrodef{CEE}[CEE]{channel estimation error}
\acrodef{CCP}[CCP]{convex-concave procedure}
\acrodef{MRT}[MRT]{ maximum-ratio transmission}
\acrodef{MM}[MM]{Minorization-Maximization}
\acrodef{PSD}[PSD]{positive semi-definite}
\acrodef{RZF}[RZF]{Regularized zero forcing}
\acrodef{CRZF}[CRZF]{Centralized regularized zero forcing}
\acrodef{RZF}[RZF]{regularized zero forcing}
\acrodef{LPZF}[LPZF]{Local Partial zero forcing}
\acrodef{LZF}[LZF]{Local zero forcing}
\acrodef{NF}[NF]{near field}
\acrodef{FF}[FF]{Far Field}
\acrodef{BD}[BD]{Beyond-diagonal}
\acrodef{OFDM}[OFDM]{orthogonal frequency division multiplexing}
\acrodef{MIMO}[MIMO]{multiple-input multiple-output}
\acrodef{MAPRT}[MAPRT]{maximum a-posteriori ratio test}
\acrodef{GLRT}[GLRT]{generalized likelihood ratio test}
\acrodef{CDF}[CDF]{Cumulative distribution function}
\acrodef{UPA}[UPA]{uniform planar array}
\acrodef{LMMSE}[LMMSE]{linear minimum mean square error}
\acrodef{SE}[SE]{spectral efficiency}
\acrodef{CNR}[CNR]{clutter to noise ratio}
\acrodef{SCNR}[SCNR]{signal to clutter and noise ratio}
\acrodef{SOCP}[SOCP]{second order cone program}
\acrodef{TTD}[TTD]{true time delay}
\acrodef{PS}[PS]{phase shifter}
\acrodef{PEB}[PEB]{position error bound}
\acrodef{JRC}[JRC]{Joint radar and communications}
\acrodef{GP}[GP]{gradient projection}
\acrodef{MO}[MO]{Manifold Optimization}
\acrodef{DEB}[DEB]{direction error bound}
\acrodef{DOA}[DOA]{direction of arrival}
\acrodef{DOD}[DOD]{direction of departure}
\begin{document}

\title{Millimeter-Wave Joint Radar and Communications With an RIS-Integrated Array
}

\author{Steven~Rivetti$^\dagger$,
Özlem~Tu$\Breve{\text{g}}$fe~Demir$^*$,
Emil~Björnson$^\dagger$, 
        Mikael~Skoglund$^\dagger$ \\
        
       {\small$^\dagger$School of Electrical Engineering and Computer Science (EECS),
        KTH Royal Institute of Technology, Sweden} \\
        
        {\small$^*$Department of Electrical-Electronics Engineering, TOBB University of Economics and Technology, Ankara, Türkiye}
        
       \thanks{
This work was supported by the SUCCESS project (FUS21-0026), funded by the Swedish Foundation for Strategic Research.}}

\maketitle

\begin{abstract}
In the context of the joint radar and communications (JRC) framework, reconfigurable intelligent surfaces (RISs) emerged as a promising technology for their ability to shape the propagation environment by adjusting their phase-shift coefficients. However, achieving perfect synchronization and effective collaboration between access points (APs) and RISs is crucial to successful operation. This paper investigates the performance of a bistatic JRC network operating in the millimeter-wave (mmWave) frequency band, where the receiving AP is equipped with an RIS-integrated array. This system simultaneously serves multiple UEs while estimating the position of a target with limited prior knowledge of its position. To achieve this, we optimize both the power allocation of the transmitted waveform and the RIS phase-shift matrix to minimize the position error bound (PEB) of the target. At the same time, we ensure that the UEs achieve an acceptable level of spectral efficiency.  The numerical results show that an RIS-integrated array, even with a small number of receiving antennas, can achieve high localization accuracy. Additionally, optimized phase-shifts significantly improve the localization accuracy in comparison to a random phase-shift configuration.

\end{abstract}


\section{Introduction}
 \ac{JRC} has emerged as a promising solution to address the spectrum scarcity challenges faced by sixth-generation (6G) wireless networks. By combining radar and communication functionalities, JRC enhances the hardware and \ac{SE}, allowing the sensing and communication tasks to cooperate rather than compete for limited spectrum resources.
 \acp{RIS} have recently gained attention as a key component of ISAC systems. They enable the detection of obstructed targets and increase the strength of the essential propagation paths. RISs consist of low-cost passive reflecting elements that are typically placed between the transmitter and the \ac{UE} or target, acting as intelligent reflectors. They apply controllable phase shifts to the incoming waveforms, effectively steering the reflected signals to desired locations \cite{alexandropoulos2021reconfigurable}.
A significant challenge in the operation of RISs is the need for precise time synchronization between the transmitting \ac{AP} and the RIS control center \cite{zhao2021cooperative} in order to dynamically update the RIS configuration. To overcome this limitation, an alternative approach involves the integration of an RIS, which functions as a smart transmitting surface, into an active antenna array \cite{huang2024hybrid}. This architecture enables high aperture gain using a transceiver with a relatively low number of active antenna elements, thereby allowing for an extension of the aperture area without deploying costly large antenna arrays. %Furthermore, it offers performance comparable to that of a large antenna array at a fraction of the cost \cite{demir2024user}. 
In \cite{demir2024user}, the authors examine the uplink performance of a cell-free massive MIMO (multiple-input multiple-output) system where each receiving AP is equipped with an RIS-integrated array. Numerical simulations demonstrate that this architecture can achieve \ac{SE} levels similar to those of a large antenna array. Similarly, \cite{raeisi2024efficient} investigates the downlink localization performance of a single-antenna transmitter equipped with a beyond-diagonal transmissive RIS, showing competitive performance compared to large antenna arrays.
To the best of our knowledge, the use of a RIS-integrated receiver within a bistatic JRC network has not yet been explored. This paper investigates JRC operations in a MIMO network operating in the millimeter-wave (mmWave) frequency band. The network comprises a transmitting AP equipped with a conventional antenna array employing a hybrid beamforming architecture, and a receiving AP with an RIS-integrated array.
We assume that there is no energy loss in the transmission between the RIS and the receiving antenna array: This can be facilitated by encasing the RIS and receiving array in an enclosed box with fully reflective sides \cite{demir2024user}. 
The proposed network wants to estimate a target's position under the assumption of limited prior knowledge of said target's position.
To this end, The system optimizes the transmit power allocation and RIS phase shifts to improve the estimation accuracy on the target's position while ensuring a minimum SE for the UEs. The SE is calculated for an \ac{LMMSE} channel estimator that accounts for the transmit AP’s hybrid beamforming architecture. Inspired by \cite{keskin2022optimal}, the transmit precoding matrix includes derivative beams, which we demonstrate to be beneficial for enhancing positioning performance. Numerical results confirm the effectiveness of the RIS-integrated architecture. 

\emph{Notation}: $\odot$ represents the Hadamard (element-wise) product. Boldface lowercase and uppercase letters denote vectors and matrices, respectively.
The trace of the matrix $\XX$ is denoted by $\Tr(\XX)$.
 $\diag(\xx)$ represents the stacking of $\xx$ on the main diagonal of a matrix.
$[\hh]_{n:m}$ represents vector elements comprised between the $n$-th and $m$-th ones, denoted by $h_n$ and $h_m$, of vector $\hh$. 
The notation $\mathcal{CN}(0,\sigma^2)$ represents the circularly symmetric complex Gaussian distribution with variance $\sigma^2$.

\vspace{-3mm}
\section{System Model}
\begin{figure}[t!]
\begin{center}
   \resizebox{0.45\textwidth}{!}{
    \input{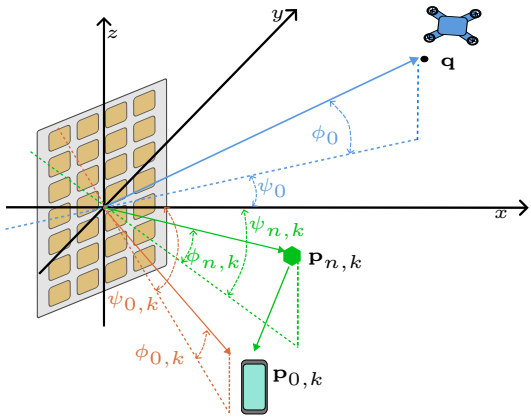}}
      \vspace{-2mm}
	 \caption{Transmit geometry of the  UPA.} \label{geo}
		\end{center}
  \vspace{-8mm}
\end{figure}

\vspace{-1mm}
We consider a massive MIMO network operating in a mmWave frequency band where a transmitting \ac{AP}, equipped with a \ac{UPA} composed of $\Mtxy\times \Mtxz =\Mtx$ antennas, serves a set $\mathcal{K}$ of $K$ single-antenna communication \acp{UE}. $\Mtxy$ and $\Mtxz$ denote the number of antennas along the $y$ and $z$ axis as shown in Fig.~\ref{geo}. We consider a bistatic sensing setup, where a receiving AP collects the radar echoes over $\tau_\textrm{s}$ symbols to estimate the position of a target.
The transmit AP adopts a hybrid fully-connected beamforming architecture and is equipped with $\Nrf$ radio frequency (RF) chains and $\Mtx$ transmit antennas.
The system employs \ac{OFDM} with $V$ subcarriers, centered around the carrier frequency $f_c$.
The receiving AP is equipped with a \ac{RIS}-integrated antenna array: here an $N$-element RIS (deployed as a UPA of $N_y\times N_z$ elements), whose phase shifts are described by $\ttheta=[\theta_1,\dots, \theta_N]^\top$, is integrated onto a conventional antenna array of $\Mrx$ receiving antennas.
We assume that $\Mrx < \Mtx$ and the receiving AP adopts a fully digital beamforming architecture.
The RIS is placed in front of the receiving AP, within its \ac{NF} \cite{demir2024user}.
The signal received by UE $k$ on subcarrier $v$ during timeslot $\iota$ is defined as
 \vspace{-1mm}
\begin{align}
    y_{k,v}[\iota]&=\underbrace{\sqrt{\rho_{k,v}}\bb_{k,v}^\hermit
\ff_{k,v}x_{k,v}[\iota] }_{\textrm{desired signal}} + \underbrace{\sum_{q \in \mathcal{Q}}\sqrt{\rho_{q,v}}\bb_{k,v}^\hermit\ff_{q,v}x_{q,v}[\iota]}_{\textrm{sensing interference}} \nonumber \\
&+\underbrace{\sum_{k' \in \mathcal{K}\setminus\{k\}}\hspace{-3mm}\sqrt{\rho_{k',v}}\bb_{k,v}^\hermit
\ff_{k',v}x_{k',v}[\iota] }_{\textrm{multi-user interference}}
 + w_{k,v}[\iota], 
\end{align}
where $\ff_{s,v}$ denotes the precoding vector employed by the AP to transmit stream $s$: the system allocates a transmission stream 
%with a matching index 
to each UE and $Q$ additional streams, whose indices are collected in the set $\mathcal{Q}$, for sensing purposes, bringing the total number of streams to $S=K+Q$.
The power allocated to each stream is denoted by  $\rho_{s,v}$.
The baseband symbol transmitted at time $\iota$ on stream $s$ and subcarrier $v$ is represented by $x_{s,v}[\iota]$, where $\mathbb{E}\{|x_{s,v}[\iota]|^2\}=1$.
The receiver noise is represented by $w_{k,v}[\iota] \sim \mathcal{CN}(0,\sigma_k^2)$.
The channel between UE $k$, placed at $\pp_{0,k}=[p_{0,k,x},p_{0,k,y},p_{0,k,z}]^\top$, and the transmitting AP is denoted by  $\bb_{k,v}$. The channel consists of a \ac{LoS} path and $C_k-1$  \ac{NLoS} paths, generated by clusters placed at  $\{\pp_{n,k}\}_{n=1}^{C_k-1}$ as demonstrated in Fig.~\ref{geo}.
Assuming that each cluster scatters a sufficient amount of rays to justify a complex Gaussian channel gain, we can model $\bb_{k,v}$ as
\vspace{-2mm}
\begin{align}
    & \bb_{k,v} =\sum_{n=0}^{C_k-1}\alpha_{n,k}\hspace{0.5mm}e^{-j2\pi v\Delta_f \tau_{n,k} }\atx(\psi_{n,k},\phi_{n,k})
\end{align}
where $\Delta_f$ is the subcarrier spacing and
\begin{align}
    \tau_{0,k}=\frac{\norm{\pp_{0,k}}}{c},\quad \tau_{n>0,k}=\frac{\norm{\pp_{n,k}} + \norm{\pp_{0,k}-\pp_{n,k}}}{c}.
\end{align}
Here, $\psi_{n,k},\phi_{n,k}$ are the angular coordinates of cluster $n$, i.e., the azimuth and elevation angles, respectively.
The $l$-th path has the complex gain $\alpha_{n,k}\sim \mathcal{CN}(0,\beta_{n,k}^2)$ and these terms are independent across $n,k$.
The large-scale fading coefficient of each cluster is denoted as $\beta_{n,k}^2$.
The spatial correlation matrix of $\bb_{k,v}$, defined as $\mathbb{E}\left\{\bb_{k,v}\bb_{k,v}^\hermit \right\}=\RR_k~\forall v$, is fixed throughout the transmission and is assumed to be known by the system. 
The phase center of the transmit AP \ac{UPA} is placed at the origin, the location of the $m$-th element is
$\pp_m=[0,m_y \frac{\lambda_c}{2},m_z\frac{\lambda_c}{2}]^\top,~m_y=0,\pm1,\cdots, \pm (\Mtxy-1)/2,~m_z=0,\pm1,\cdots, \pm (\Mtxz-1)/2$, where $\lambda_c$ is the carrier wavelength, and $\Mtxy$ and $\Mtxz$ are assumed to be odd.%without loss of generality.
Under the assumption $\frac{c}{f_c} \approx \frac{c}{f_v}~\forall v$, the transmit AP's array response vector $\atx(\psi,\phi)$ is given by
\begin{align}
    &\atx(\psi,\phi) = \left[ e^{j\mathbf{k}^\top(\psi,\phi)\pp_1},\cdots,e^{j\mathbf{k}^\top(\psi,\phi)\pp_{\Mtx}} \right]^\top,\\
    &\mathbf{k}(\psi,\phi) = \frac{2\pi f_c}{c} [\cos(\phi)\cos(\psi),  \cos(\phi)\sin (\psi), \sin(\phi)]^\top. 
\end{align}
The precoding matrix employed by the AP is defined as 
$\FF_v= \WW\DD_v =[\ff_{1,v},\cdots,\ff_{S,v}]\in \mathbb{C}^{\Mtx\times S}$,
 where $\WW\in \mathbb{C}^{\Mtx \times \Nrf}$ is implemented through analog phase shifters and $\DD_v\in\mathbb{C}^{\Nrf \times S}$ is the digital precoding matrix at subcarrier $v$.

\subsection{Sensing Observation Model}
% The transmitted waveform on subcarrier $v$ can be described by 
% \begin{align}
%     \ssb_v[\tau] = \FF_v\XX_v[\tau]\sqrt{\rrho_v}
% \end{align}
The echo signal observed by the receiving AP 
at subcarrier $v$ during timeslot $\iota$ is given by
\vspace{-2mm}
\begin{align}\label{sensing obs}
    &\yy_v[\iota] =\sum_{s=1}^S \underbrace{\HH_v\ff_{s,v}x_{s,v}[\iota]\sqrt{\rho_{s,v}}}_{\overline{\yy}_{s,v}[\iota]} +\ww_v[\iota] ,
\end{align}
%Where $\XX_v[\tau] = \diag(\boldsymbol{1}_Q^\top,x_{1,v}[\tau],\dots,x_{K,v}[\tau])$ and $\rrho_v=[\rho_{v,1},\dots,\rho_{v,S}]^\top$. 
where the two-way radar channel is defined as 
\begin{align}
    \HH_v =\alpha_\rmt e^{-j2\pi v \Delta_f\tau_\rmt } 
 \TT_v\TTheta
 \aris(\psi_\rmt,\phi_\rmt) \atx^\top(\psi_0,\phi_0). 
\end{align}
We do not consider the link between the transmitting and receiving APs as it does not give any information about the
estimation of the target’s position, as the position and
orientation of the APs is assumed to be
known\cite{laas2024optimal}
%, and clutter has been removed .
The round-trip delay is defined as $\tau_\rmt= \frac{\norm{\qq}}{c} + \frac{\norm{\ee-\qq}}{c}$, where $\qq=[q_{x},q_{y},q_{z}]^\top$ and $\ee=[e_{x},e_{y},e_{z}]^\top$ denote the target and RIS locations, respectively.
The complex channel gain associated with this path, which also takes into account the target's \ac{RCS}, is represented by $\alpha_\rmt\sim \mathcal{CN}(0,\beta_\rmt^2 \delta^2)$:  here $\beta_\rmt^2$ accounts for the pathloss while $\delta^2$ represent the target's RCS power.
The location of the $n$-th RIS transmissive element with respect to (w.r.t.) the RIS phase center is given by $\rr_n=[0,n_y \frac{\lambda_c}{2},n_z\frac{\lambda_c}{2}]^\top,~n_y=0,\pm1,\cdots, \pm (N_y-1)/2,~n_z=0,\pm1,\cdots, \pm (N_z-1)/2$.
The RIS's array response vector $\aris(\psi,\phi)$ is given by
$    \aris(\psi,\phi) = \left[ e^{j\mathbf{k}^\top(\psi,\phi)\rr_1},\cdots,e^{j\mathbf{k}^\top(\psi,\phi)\rr_{N}} \right]^\top$.
The target's angular coordinates w.r.t. both APs are computed as 
\vspace{-2mm}
\begin{align}
     &\psi_0=\textrm{atan2}(q_{\rm y},q_{\rm x}),~~ \phi_0 =  \textrm{atan}\left(\frac{q_{\rm z}}{\sqrt{q_{\rm x}^2+q_{\rm y}^2}}\right)\\
      &\psi_\rmt=\textrm{atan2}(e_{\rm y}-q_{\rm y},e_{\rm x}-q_{\rm x})\\
    & \phi_\rmt=  \textrm{atan}\left(\frac{e_{\rm z}- q_{\rm z}}{\sqrt{({e_{\rm x}- q_{\rm x})^2+({e_{\rm y}- q_{\rm y})^2} }}}\right).
\end{align}
The RIS's phase-shift configuration is described by $\TTheta = \diag(\ttheta)$ and 
$\ww_v[\iota] \sim \mathcal{CN}({\bf 0},\sigma^2\mathbf{I}_{\Mrx})$ is the receiver noise at timeslot $\iota$ and subcarrier $v$.
We denote by $\TT_v \in \mathbb{C}^{\Mrx \times N}$ the NF channel connecting the receiving antennas to the transmitting RIS. This channel is composed by a LoS path component and $C_R-1$ NLoS ones.
The \ac{NLoS} paths originate by the reflection off the sides of the enclosed box and are modelled by scatterers placed at $\{\zz_i\}_{i=1}^{C_R-1}$.
%originated by the reflections from the side of the enclosed box containing both the RIS and the receiving AP.
Each element of $\TT_v$ is then given by 
\vspace{-2mm}
\begin{align}
    &[\TT_v]_{m,n}=\zeta_{m,n,0}  e^{-j\frac{2\pi f_c}{c}d_{m,n,0}} e^{-j2\pi\Delta_fv\frac{d_{m,n,0}}{c}} + \\
    &\sum_{i=1}^{C_R-1}\kappa_{m,n,i} \zeta_{m,n,i}  e^{-j\frac{2\pi f_c}{c}d_{m,n,i}} e^{-j2\pi\Delta_fv\frac{d_{m,n,i}}{c}}
\end{align}
where $\zeta_{m,n,0}=c/(4\pi f_c d_{m,n,0})$ and $\zeta_{m,n,i}\sim \mathcal{CN}(0,c^2/(4\pi f_c d_{m,n,i})^2) $, for $i=1,\ldots,C_R-1$ represent the path loss and 
\vspace{-2mm}
\begin{align}
    &d_{m,n,0}=\norm{\rr_n-\overline{\pp}_m},\\
    &d_{m,n,i>0}=
    \norm{\rr_n-\zz_i} + \norm{\zz_i-\overline{\pp}_m},
\end{align}
where $\overline{\pp}_m$ is the location vector of the $m$-th antenna of the receive AP. The coefficient $\kappa_{m,n,i}$ ensures that the power of
The NLoS component is adjusted such that the norm of each
column of $\TT_v$ is equal to one. This 
ensures that
the energy collected by each RIS element is distributed across all AP antennas \cite{demir2024user}.
Furthermore, this channel is assumed to be constant due to the non-varying propagation environment between RIS and AP.

\subsection{CRB-based Performance Metric}
We assume that a two-step positioning approach is implemented, consisting of the estimation of an unknown channel parameter followed by the estimation of a target's position. The channel parameters to be estimated are collected in the vector 
 $   \boldsymbol{\eta}= [\psi_0,\psi_\rmt,\phi_0,\phi_\rmt,\tau_\rmt,|\alpha_\rmt|,\angle\alpha_\rmt]$.
Under the assumption that the correlation between symbols belonging to different streams on the same subcarrier or between symbols belonging to different subcarriers is zero, the \ac{FIM} in the channel domain is computed as
\begin{align}
   \mathbf{J} = \frac{2}{\sigma^2} \sum_{\iota=1}^{\tau_\textrm{s}}\sum_{v=1}^V\sum_{s=1}^S  \Re \Bigg\{ \bigg( \frac{\partial \overline{\yy}_{s,v}[\iota]}{\partial  \boldsymbol{\eta}} \bigg)^\hermit  \bigg( \frac{\partial \overline{\yy}_{s,v}[\iota]}{\partial  \boldsymbol{\eta}} \bigg)  \Bigg\}.
\end{align}
We can now define a vector of unknowns in the location domain as $\overline{\eeta}=[\qq^\top,|\alpha_\rmt|,\angle\alpha_\rmt] $.
 The FIM w.r.t. $\boldsymbol{\overline{\eta}}$, denoted by $\overline{\JJ}$, is obtained by means of a matrix transformation as
 $\overline{\JJ} = \boldsymbol{\Xi}^\top \JJ  \boldsymbol{\Xi}$, % + \JJ^\textrm{prior},
 where  $[\boldsymbol{\Xi}]_{i,j}=\partial[\boldsymbol{\eta}]_i/\partial 
 [\boldsymbol{\overline{\eta}}]_j$ .
 %Here $\JJ^\textrm{prior}$ represent's the FIM obtained from the system's prior knowledge: give that the only random parameter in $\overline{\eeta}$ is $\Delta_t$, then $[\JJ^\textrm{prior}]_{4,4}=\sigma_\textrm{clk}^{-2}$ and zero everywhere else.
%is defined as 
% \begin{align}
%     &\boldsymbol{\Xi}= \frac{\partial \boldsymbol{\eta}^\top}{\partial \boldsymbol{\boldsymbol{\overline{\eta}}}}=\\
%     &\begin{bmatrix}
%         \partial \psi_0/\partial q_{x}              &    \partial \psi_\rmt/\partial q_{x}             & \dots & \partial \Im\{\alpha_\rmt\}/\partial q_{x}             \\
%         \partial \psi_0/\partial q_{y}              &    \partial \psi_\rmt/\partial q_{y}             & \dots & \partial \Im\{\alpha_\rmt\}/\partial q_{y}              \\
%         \partial \psi_0/\partial q_{z}              &    \partial \psi_\rmt/\partial q_{z}             & \dots & \partial \Im\{\alpha_\rmt\}/\partial q_{z}              \\
%         \partial \psi_0/\partial \Re\{\alpha_\rmt\} &    \partial \psi_\rmt/\partial \Re\{\alpha_\rmt\}& \dots & \partial \Im\{\alpha_\rmt\}/\partial \Re\{\alpha_\rmt\} \\
%         \partial \psi_0/\partial \Im\{\alpha_\rmt\} &    \partial \psi_\rmt/\partial \Im\{\alpha_\rmt\}& \dots & \partial \Im\{\alpha_\rmt\}/\partial \Im\{\alpha_\rmt\} \\
%     \end{bmatrix}.\nonumber
% \end{align}
 The \ac{CRB} on the target's position, hereby referred to as the \ac{PEB}, can then be defined as 
\begin{align}
        \textrm{PEB}(\TTheta,\{\rho_{s,v}\},\overline{\eeta})=\sqrt{ \sum_{n=1}^3\left[\overline{\JJ}^{-1}\right]_{n,n} }.
\end{align}

 \subsection{Channel Estimation}
The first $\tau_\textrm{p}$ samples of each coherence block are used to estimate the UE channels. We consider a set of $\tau_{\rm{p}}$ mutually orthogonal pilot sequences $\zzeta_1,\dots,\zzeta_{ \tau_{\rm{p}} } \in \mathbb{C}^{\tau_{\rm{p}}} $. The norm square of each sequence is $\tau_{\rm{p}}$.
The observation at subcarrier $v$ during the uplink pilot transmission phase is described as
\begin{align}
    &\YY_v^{\rm p} = \sum_{k=1}^K\sqrt{\mu_{k}}\GG^\hermit\bb_{k,v}\zzeta_{t_k}^\top + \GG^\hermit\NN_v    \in \mathbb{C} ^{\Nrf \times \tau_\textrm{p}},
\end{align}
where $\mu_{k}$ is the transmit power of UE $ k$, $t_k$ is the index of the pilot sequence assigned to UE $k$, and $\NN_v$, whose entries are independent and identically distributed (i.i.d.) $\mathcal{CN}(0,\sigma^2)$, is the receiver noise.
Due to the adopted hybrid beamforming architecture, the whole $\Mtx$-dimensional uplink pilot signal cannot be fully observed \cite{haghighatshoar2016enhancing}.
The system can only observe an $\Nrf$-dimensional projection of said signal. This is practically implemented by the presence of an analog combining matrix $\GG \in \mathbb{C}^{\Mtx \times \Nrf}$. 
Assuming that $\Nrf$ is a multiple of $K$ and by denoting as $\uu_{k,l}$ the eigenvector associated to the $l$-th largest eigenvalue of $\RR_k$, the analog combiner can be defined as 
\begin{align}
    \GG=[\uu_{1,1},\dots,\uu_{1,L},\dots,\uu_{K,1},\dots,\uu_{K,L}]
\end{align}
with $L=\Nrf/K$.
%Thanks to the matrix property $\vecc(\AAb\BB\CC)=(\CC^\top\otimes\AAb)\vecc(\BB)$, the vectorized version of $\YY_v^{\rm p}$, which is denoted by $\yy_v^{\rm p}$, is given by
% \begin{align}
%     &\yy_v^{\rm p}=\sum_{k=1}^K\underbrace{\sqrt{\mu_{k}} ( \zzeta_k \otimes \GG_\rf^\hermit)}_{\AAb_k\in \mathbb{C}^{\Nrf\tau_\textrm{p}\times \Mtx}}\bb_{k,v} + \widetilde{\nn}_v \in \mathbb{C}^{\Nrf\tau_\textrm{p}}
% \end{align}
% where $\widetilde{\nn}_v=\vecc(\widetilde{\NN}_v)\sim \mathcal{CN}(0,\ZZ)$ , with $\ZZ = \sigma^2(\mathbf{I}_{\tau_\textrm{p}} \otimes\GG_\rf^\hermit\GG_\rf)$.
Assuming there is no pilot contamination, i.e., $\tau_{\rm p}\geq K$, the UEs' channels can be estimated through an \ac{LMMSE} estimation procedure as 
\begin{align}\label{ch est}
    \widehat{\bb}_{k,v}= \sqrt{\mu_k\tau_\textrm{p}}\RR_k\GG\Big( \mu_k\tau_\textrm{p}  \GG^\hermit \RR_k \GG +\sigma^2\GG^\hermit\GG    \Big)^{-1} \YY_v^{\rm p}\zzeta_{t_k}^*.
\end{align}
% where $\mathcal{P}_k$ is the set of UEs sharing the pilot $\zzeta_{t_k}$, defined as 
% \begin{align}
%     \mathcal{P}_k = \{k' : t_{k'}=t_k,~\forall~k'\in \mathcal{K}\} \subset \mathcal{K}.
% \end{align}
\subsection{Communication Precoding Matrix}
We assume that only $N_\textrm{RF}^\textrm{dl}=K+Q$ of the available RF chains are used during downlink data transmission.
Given the sparse nature of  $\bb_{k,v}$, the communication analog precoding matrix is defined as
\begin{align}
    \WW^{\mathcal{K}}=[\tilde{\uu}_{1,1},\dots,\tilde{\uu}_{K,1}]\in \mathbb{C}^{\Mtx \times K}
\end{align}
where $\tilde{\uu}_{k,1}$ is the best analog approximation of $\uu_{k,1}$, obtained through gradient projection (GP) \cite[Alg.~1]{tranter2017fast}.
As for the digital communication precoding matrix, we implement \ac{RZF}. We define the effective channel vector $\widehat{\overline{\bb}}_{k,v}=\WW^\hermit \widehat{\bb}_{k,v} $, then the communication digital precoding matrix is 
$\DD_v^\mathcal{K}=[\dd_{1,v},\dots,\dd_{K,v}]$, where
$\dd_{k,v}=\overline{\dd}_{k,v}/\norm{\WW\overline{\dd}_{k,v}}$ in which
\begin{align}
    \overline{\dd}_{k,v}=\left( \sum_{k=1}^K \mu_k\widehat{\overline{\bb}}_{k,v}\widehat{\overline{\bb}}_{k,v}^\hermit + \sigma^2 \mathbf{I}_{N_\textrm{RF}^\textrm{dl}}\right)^{-1}\hspace{-3mm}\sqrt{\mu_k}\hspace{0.5mm}\widehat{\overline{\bb}}_{k,v}.
\end{align}
The communication performance metric of \ac{UE} $k$ is the \ac{SE} on each subcarrier. Assuming that the UEs know $\mathbb{E}\{\bb_{k,v}^\hermit \ff_{k,v}\}$, the use-and-then-forget bound allows us to achieve the SE
 \begin{align}
     {\rm{SE}}_{k,v} = \frac{\tau_\textrm{c}-\tau_\textrm{p}}{\tau_\textrm{c}}\log_2(1+{\rm{SINR}}_{k,v}),
 \end{align}
 where $\tau_\textrm{c}$ is the channel coherence length in terms of samples.
The effective \ac{SINR} of UE $k$ at subcarrier $v$ is given in \eqref{multipage} on top of the next page, with $\widetilde{\bb}_{k,v}=\bb_{k,v}-\widehat{\bb}_{k,v}$ denoting the channel estimation error. As we will show later, we select the sensing precoding vectors in the null-space of the UE channel estimates.
%----------------------------------------
\begin{figure*}[t!]
  \begin{align}\label{multipage}
   &{\rm{SINR}}_{k,v} = \frac{\rho_{k,v}\left \vert\mathbb{E}\{\bb_{k,v}^\hermit \ff_{k,v}\}\right\vert^2}{\sum_{j=1}^{K} \rho_{j,v} \mathbb{E}\{ |\bb_{k,v}^\hermit \ff_{j,v}|^2 \} - \rho_{k,v}\left \vert\mathbb{E}\{\bb_{k,v}^\hermit \ff_{k,v}\}\right\vert^2 
   +
   \sum_{q \in \mathcal{Q}}\rho_{q,v}\mathbb{E}\{|\widetilde{\bb}_{k,v}^\hermit \ff_{q,v}|^2
   \}+\sigma^2_k} =\frac{f_{k,v}\left(\{\rho_{s,v}\}\right)}{g_{k,v}(\{\rho_{s,v}\})}
\end{align}  
\hrulefill
\vspace{-6mm}
\end{figure*}
%---------------------------------------
The terms in the SINR are computed as outlined in 
\cite{behdad2024multi}.

\subsection{Sensing Precoding Matrix}
The system has some prior knowledge about the unknown position, that is, $\overline{\eeta} \in \mathcal{U}$. 
This in turn defines a rectangular area in the $(\psi_0,\phi_0)$ domain, denoted by $\mathcal{A}$: the role of the sensing precoding matrix is to cover this area in a way that optimizes the accuracy on the target's position estimation.
We compare three different sensing analog precoding matrices.
\subsubsection{Sum-derivative matrix}
We discretize $\mathcal{A}$ in  $\{(\psi_n,\phi_n)\}_{n=1}^{N_\mathcal{A}}$: inspired by \cite{keskin2022optimal}, the sensing analog precoding matrix $\WW^{\mathcal{Q}}\in \mathbb{C}^{\Mtx \times Q}$ is set to $\WW^{\textrm{SD}}=\left[\WW^{\textrm{Dir}}, \WW^{\textrm{Az}},\WW^{\textrm{El}}\right]\mathbb{C}^{\Mtx \times 3N_\mathcal{A}} $, where
\begin{align}
    &\WW^{\textrm{Dir}}=[\atx(\psi_1,\phi_1),\dots,\atx(\psi_{N_\mathcal{A}},\phi_{N_\mathcal{A}})]\\
    &\WW^{\textrm{Az}}=[\atxdttilde(\psi_1,\phi_1),\dots,\atxdttilde(\psi_{N_\mathcal{A}},\phi_{N_\mathcal{A}})]\\
    &\WW^{\textrm{El}}=[\atxddttilde(\psi_1,\phi_1),\dots,\atxddttilde(\psi_{N_\mathcal{A}},\phi_{N_\mathcal{A}})]\\
&[\atxdt(\psi,\phi)]_m\hspace{-1mm}=\left(j\frac{\partial\mathbf{k}^\top(\psi,\phi)\pp_{m}}{\partial \psi}\right)[\atx(\psi,\phi)]_{m}\\
&[\atxddt(\psi,\phi)]_m=\left(j\frac{\partial \mathbf{k}^\top(\psi,\phi)\pp_{m}}{\partial \phi}\right)[\atx(\psi,\phi)]_{m}.
\end{align}
By $\atxdttilde(\psi,\phi),\atxddttilde(\psi,\phi)$ we denote the analog approximations of $\atxdt(\psi,\phi),\atxddt(\psi,\phi)$ once again obtained through GP \cite[Alg.~1]{tranter2017fast}.
\subsubsection{Sum-only matrix}
To assess the impact of derivative beams, we now employ only the directional beams used in $\WW^{\textrm{SD}}$, that is
$\WW^{\mathcal{Q}}=\WW^{\textrm{Dir}}$.
%The RF chains employed by the sensing analog codebook are now only $Q=N_\mathcal{A}$.
\subsubsection{Extended sum-only matrix}
By adopting the same number of RF chains as $\WW^\textrm{SD}$, We now discretize $\mathcal{A}$ in $\{(\psi_n,\phi_n)\}_{n=1}^{Q} $, setting $\WW^{\mathcal{Q}}$ to $\WW^{\textrm{SE}}$, chosen as 
\begin{align}
\WW^{\textrm{SE}}=\hspace{-0.5mm} \left[\atx(\psi_1,\phi_1),\dots,\atx(\psi_{Q},\phi_{Q})\right].
\end{align}
Regardless of the chosen analog precoding matrix, the role of the digital sensing precoding matrix $\DD_v^\mathcal{Q}=[\dd_{1,v},\dots,\dd_{Q,v}]$  is  nulling the destructive interference between sensing and communication beams: We thus chose $\dd_{q,v}=\overline{\dd}_{q,v}/\norm{\WW\overline{\dd}_{q,v}}$ as a zero-forcing precoder, that is 
\begin{align}
\overline{\dd}_{q,v}=\left(\mathbf{I}_{N_\textrm{RF}^\textrm{dl}} - \UU_v\UU_v^\hermit\right)\WW^\hermit\ww_{q}
\end{align}
where $\UU_v$ is a unitary matrix with orthogonal columns spanning the column space of $\widehat{\overline{\BB}}_{v}=[\widehat{\overline{\bb}}_{1,v},\dots, \widehat{\overline{\bb}}_{K,v}]$.
Eventually, we can define the complete precoding matrices as 
\begin{align}
    &\WW=[\WW^\mathcal{K},\WW^\mathcal{Q}] \in \mathbb{C}^{\Mtx \times N_\textrm{RF}^\textrm{dl}}\\
    &\DD_v=[\DD_v^\mathcal{K},\DD_v^\mathcal{Q}] \in \mathbb{C}^{N_\textrm{RF}^\textrm{dl} \times (K+Q)}.
\end{align}
\section{Sensing-centric signal design}
As mentioned in the previous section, the system has  prior knowledge about the system's position unknowns, we thus resort to a worst-case PEB minimization strategy, more specifically
\begin{align}\label{FIM trans}
     \underset{\TTheta,\{\rho_{s,v}\}}{\mathrm{minimize}}~\underset{\overline{\boldsymbol{\eta}}\in \mathcal{U}}{\mathrm{max}}&~ \textrm{PEB}(\TTheta,\{\rho_{s,v}\},\overline{\eeta})\\ 
     \text{subject to} ~& \text{SE}_{k,v} \geq \gamma_{k,v},\quad \forall k,v, \label{SE og} \\
     & \sum_{v=1}^V \mathbb{E}\left\{\norm{\FF_{v}\PP_{v}\xx_v[\iota]}^2\right\} \leq P_\tx\label{power og},\\
     %\Tr\left(\FF \PP \XX \FF^\hermit\right) = P\tau_\textrm{s} \\
     & |\theta_n|=1, \forall n, \label{theta og}
\end{align}
where 
%$\FF=\blkdiag(\FF_1,\dots,\FF_V)$, 
$\PP_v =\diag(\sqrt{\rho_{1,v}},\dots,\sqrt{\rho_{S,v}})$, $\xx_v[\iota]=[x_{1,v}[\iota],\dots,x_{S,v}[\iota]]^\top$ and $\gamma_{k,v}$ is the SE that is to be guaranteed to the UEs. $P_\tx$ is the transmit power available at the transmit AP. 
%\eqref{cnr constr} represent a constraint on the average \ac{CNR}, we assume that $\mathbb{E}\left\{ \hh_{\textrm{cl},v}\hh_{\textrm{cl},v}^\hermit \right\}=\CC_v$ is known by the system.
We can then discretize $\mathcal{U}$ into a grid of $N_\mathcal{U}$ points $\{\overline{\eeta}_n\}_{n=1}^{N_\mathcal{U}}$and recast the previous problem into its epigraph form as 
\begin{subequations}
  \begin{align}\label{problem og}
    \underset{\TTheta,\{\rho_{s,v}\},\{u_{n,b}\},r}{\mathrm{minimize}}~r\\ 
     \text{subject to} ~& \begin{bmatrix}
        \hspace{0.5mm}\overline{\JJ}(\TTheta,\{\rho_{s,v}\},\overline{\eeta}_n) & \mathbf{i}_b \\
        \mathbf{i}_b^\top & u_{n,b}
    \end{bmatrix} \succeq 0,~b=1,2,3,~\nonumber\\
    &n=1,\dots,N_\mathcal{U}, \label{schur constr}\\
    &\sum_{b=1}^3 u_{n,b}  \leq r , \quad  n=1,\dots, N_\mathcal{U}, \label{u constr}\\
& \eqref{SE og},~\eqref{power og},~\eqref{theta og},
\end{align}  
\end{subequations}
where $\mathbf{i}_b$ represents the $b$-th column of the identity matrix and
 $\overline{\JJ}(\TTheta,\{\rho_{s,v}\},\overline{\eeta}_n)$ represents the FIM $\overline{\JJ}$ evaluated at $\eeta=\eeta_n$, while $\{u_{n,b}\},r$ are auxiliary variables.
This problem is nonconvex due to the complicated coupling between the optimization variables, so we will split it into two subproblems and solve them iteratively.

\subsection{Power Allocation Optimization}
One can show that each term of $\JJ$ can be written as 
\begin{align}
    \left[ \JJ \right]_{i,j}=&\frac{2}{\sigma^2} \sum_{\iota=1}^{\tau_\textrm{s}} \sum_{v=1}^V\sum_{s=1}^S  \Re \left\{\rho_{s,v}|x_{s,v}[\iota]|^2\ff_{s,v}^\hermit \frac{\partial \HH_{v}^\hermit}{\partial  \eta_i} \frac{\partial \HH_{v}}{\partial  \eta_j}\ff_{s,v}\right\}\nonumber\\
    =&\frac{2}{\sigma^2}\sum_{v=1}^V  \Re \left\{ \Tr\left(\underbrace{\FF_v\PP_v^2\XX_v\FF_v^\hermit}_ {\SSb_v} \frac{\partial \HH_{v}^\hermit}{\partial  \eta_i} \frac{\partial \HH_{v}}{\partial  \eta_j}\right)\right\}
    % =&\frac{2}{\sigma^2}\Re\left\{ \Tr\left(\FF \PP \XX \FF^\hermit \GG_{i,j}\right)\right\}
\end{align}
where $\XX_v=\sum_{\iota=1}^{\tau_\textrm{s}}\diag(|x_{1,v}[\iota]|^2,\dots,|x_{S,v}[\iota]|^2)$.
Since $\overline{\JJ}$ is obtained from $\JJ$ through a linear transformation and each term of $\JJ$ has a linear dependency on $\{\rho_{s,v}\}$, $\overline{\JJ}$ is affine in $\{\rho_{s,v}\}$.
We can rewrite \eqref{SE og} as a minimum SINR constraint, which is convex as the latter exhibits a linear dependency from $\{\rho_{s,v}\}$. The power allocation minimization subproblem then becomes  
\vspace{-3mm}
\begin{subequations}\label{sub power}
  \begin{align}
  \underset{ \{\rho_{s,v}\}\{u_{n,b}\},r}{\mathrm{minimize}}~r\\ 
     \text{subject to} ~& \begin{bmatrix}
\hspace{0.5mm}\overline{\JJ}(\TTheta,\{\rho_{s,v}\},\overline{\eeta}_n) & \mathbf{i}_b \\
        \mathbf{i}_b^\top & u_{n,b}
    \end{bmatrix} \succeq 0, \nonumber\\
    &~b=1,2,3,~n=1,\dots,N_\mathcal{U},\\
    &f_{k,v}\left(\{\rho_{s,v}\}\right) \geq g_{k,v}\left(\{\rho_{s,v}\}\right) \widetilde{\gamma}_{k,v} ,~\forall k,v,\\
    &\eqref{u constr}\\
    &\Tr\left(\FF\PP\FF^\hermit\right) \leq P_\tx
\end{align}  
\end{subequations}
where $\widetilde{\gamma}_{k,v}=2^{\frac{\tau_\textrm{c}}{\tau_\textrm{c}-\tau_\textrm{p}}\gamma_{k,v}}-1$, $\FF=\blkdiag(\FF_1,\dots,\FF_V)$ and $\PP=\blkdiag(\PP_1^2,\dots,\PP_V^2)$. This problem is convex and can be solved using general-purpose convex optimization solvers, like CVX.
\vspace{-1mm}
\subsection{RIS Optimization}
We can write the two-way radar channel as $\HH_v=\TT_v\TTheta\overline{\HH}_v$, where
$    \overline{\HH}_v=\alpha_\rmt e^{-j2\pi v\Delta_f\tau_\rmt } 
 \aris(\psi_\rmt,\phi_\rmt) \atx^\top(\psi_0,\phi_0)$. 
 By using the property $\Tr(\AAb\XX\BB\XX^\hermit)=\vecc(\XX)^\hermit(\BB^\top \otimes \AAb)\vecc(\XX)$ and $\text{\normalfont vec}^\hermit(\diag(\xx)) ( \AAb \otimes \BB ) \text{\normalfont vec}(\diag(\xx)) = \xx^\hermit (\AAb \odot \BB ) \xx$ from \cite{horn2012matrix}, one can show that each element of $\JJ$ can be written as
 \vspace{-1mm}
\begin{align}
     \hspace{-1mm}\left[ \JJ \right]_{i,j}\hspace{-0.5mm}&=\frac{2}{\sigma^2} \Re \left\{ \ttheta^\hermit \left( \sum_{v=1}^V \left(\dot{\overline{\HH}}_{v,j}\SSb_v \dot{\overline{\HH}}_{v,i}^\hermit\right)^\top \hspace{-2mm}\odot \TT_v^\hermit\TT_v\right)\ttheta\right\} 
     %\frac{2}{\sigma^2}\sum_{v=1}^V \Re \left\{ \Tr\left(\FF_v \PP_v \XX_v \FF_v^\hermit  \dot{\overline{\HH}}_{v,i}^\hermit \TTheta^\hermit \TT_v^\hermit \TT_v \TTheta \dot{\overline{\HH}}_{v,j} \right)\right\}\nonumber\\
\end{align}
where $\dot{\overline{\HH}}_{v,i}^\hermit =\partial \overline{\HH}_{v}/\partial  \eta_i$.
By adopting the variable change $\boldsymbol{\Gamma}=\ttheta\ttheta^\hermit$, we can write the RIS optimization problem as  
\vspace{-2mm}
\begin{subequations}\label{sub ris}
  \begin{align}
      \underset{\boldsymbol{\Gamma},\{u_{n,b}\},r}{\mathrm{minimize}}&~r \\ 
     \text{subject to}~& \begin{bmatrix}    \hspace{0.5mm}\overline{\JJ}(\boldsymbol{\Gamma},\{\rho_{s,v}\},\overline{\eeta}_n) & \mathbf{i}_b \\
        \mathbf{i}_b^\top & u_{n,b}
    \end{bmatrix} \succeq 0,~b=1,2,3,~\nonumber\\
    &n=1,\dots,N_\mathcal{U}, \\
     &\eqref{u constr}\\
    & \left[\boldsymbol{\Gamma}\right]_{n,n} =1,\quad n=1,\ldots, N_\mathcal{U},\\
    &\textrm{rank}(\boldsymbol{\Gamma})=1.
\end{align}  
\end{subequations}
This problem can be relaxed by removing the rank 1 constraint and obtaining a convex semidefinite program; the RIS phase shifts can then be recovered through Gaussian randomization \cite{luo2004multivariate}.
The optimization procedure is outlined in Algorithm~1.

\begin{algorithm} [t!]
  \caption{PEB optimization algorithm}
  \begin{algorithmic}[1]
    \STATE \textbf{Channel estimation} 
        \STATE Obtain the uplink pilot observations $\YY_v^\textrm{p}$ trough the analog combiner $\GG$.
        \STATE Obtain the LMMSE channel estimates as in \eqref{ch est}
    \STATE \textbf{Downlink transmission optimization}
    \STATE Randomly generate $\TTheta^{(0)}$, set $t=0$
    \REPEAT
        \STATE Compute $\FF_v$ 
        \STATE Obtain $\{\rho_{s,v}^{(t+1)}\}$ by solving \eqref{sub power} 
         \STATE Obtain $\TTheta^{(t+1)}$ by solving \eqref{sub ris}, $t \leftarrow t+1$
\UNTIL{$\left\vert\textrm{PEB}^{(t)} - \textrm{PEB}^{(t+1)}\right\vert$} is sufficiently small
  \end{algorithmic}
\end{algorithm}
\vspace{-2mm}
\section{Numerical Results}
\vspace{-1mm}
In this section, the numerical simulation results are provided: the transmitting AP is located in the origin, whereas the receiving RIS is located at $\ee=[100, 13,10]^\top$ [m]. The receiving AP's antenna array, configured as a uniform linear array alongside the $y$ axis, is located at $[100.3, 13,10]^\top$ [m]. The target is located at $\qq=[50,50,5]^\top$ [m] whereas the $K=5$ UEs, to which the system guarantees a minimum SE of $\gamma = 0.8$ [bps/Hz], are uniformly distributed in a $40 \times 20$ [m] area, having its bottom-left corner in $[60, 30,-9]$[m].
Unless otherwise specified, $\Mtxy=\Mtxz=N_y=N_z=7$ and $\Mrx=4$ whereas $\Nrf=20$.
 We consider $V=32$ subcarriers around a carrier frequency of $f_c=30$\,GHz, occupying a bandwidth of $B=40$\,MHz \cite{fascista2019millimeter}. the channel coherence time is $\tau_\textrm{c}=15$ timeslots, $\tau_\textrm{p}=5$ of which are dedicated to channel estimation and $\tau_\textrm{s}=10$ are used for sensing and communication.
The power budget for each transmit AP is $35$\,dBm, and the noise power is equal to $-204+10\log_{10}(B)+F_\textrm{noise}$\,dBW, where 
$F_\textrm{noise}=4$\,dB is the noise figure. The channel's path loss follows the 3GPP urban microcell model defined in \cite{3gpp2016technical}.
Unless otherwise specified, the uncertainty region $\mathcal{U}$ is $20 \times 20 \times 10$ [m] centered at the target location and is discretized in $N_\mathcal{U}=10$ points and $N_\mathcal{A}=3$.
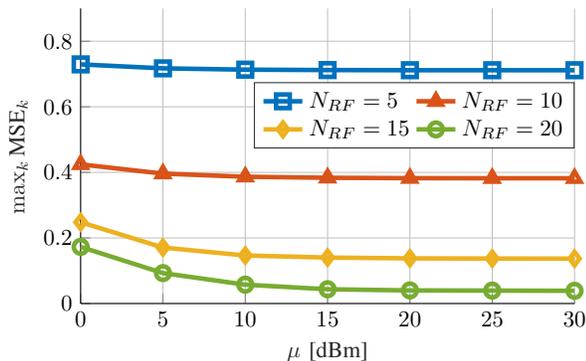
\begin{figure}[t!]
\begin{center}
   \resizebox{0.44\textwidth}{!}{
    % This file was created by matlab2tikz.
%
%The latest updates can be retrieved from
%  http://www.mathworks.com/matlabcentral/fileexchange/22022-matlab2tikz-matlab2tikz
%where you can also make suggestions and rate matlab2tikz.
%
\definecolor{mycolor1}{rgb}{0.00000,0.44700,0.74100}%
\definecolor{mycolor2}{rgb}{0.85000,0.32500,0.09800}%
\definecolor{mycolor3}{rgb}{0.92900,0.69400,0.12500}%
\definecolor{mycolor5}{rgb}{0.46600,0.67400,0.18800}%
\begin{tikzpicture}

\begin{axis}[%
width=0.42\textwidth,
height=1.793in,
at={(0.758in,0.499in)},
scale only axis,
xmin=0,
xmax=30,
xlabel style={font=\color{white!15!black}},
xlabel={$\mu$ [dBm]},
ymin=0,
ymax=0.9,
legend columns=2,
ylabel style={font=\color{white!15!black}},
ylabel={$\max_k \textrm{MSE}_k $},
axis background/.style={fill=white},
axis x line*=bottom,
axis y line*=left,
xmajorgrids,
ymajorgrids,
legend style={legend cell align=left,,at={(1,0.75)}, align=left, draw=white!15!black}
]
\addplot [color=mycolor1, line width=2.0pt, mark=square, mark options={solid, mycolor1, mark size=3pt}]
  table[row sep=crcr]{%
0	0.729427331672715\\
5	0.717380181224967\\
10	0.713342579732888\\
15	0.712106755598629\\
20	0.711606833528276\\
25	0.71160916790107\\
30	0.71155042203605\\
};
\addlegendentry{$N_{RF}=5$}

\addplot [color=mycolor2, line width=2.0pt, mark=triangle, mark options={solid, mycolor2, mark size=3pt}]
  table[row sep=crcr]{%
0	0.424388558613521\\
5	0.396185784738083\\
10	0.386643952572064\\
15	0.383520941595051\\
20	0.382356926043822\\
25	0.382065470299827\\
30	0.382011881917103\\
};
\addlegendentry{$N_{RF}=10$}
\addplot [color=mycolor3, line width=2.0pt, mark=diamond, mark options={solid, mycolor3, mark size=3pt}]
  table[row sep=crcr]{%
0	0.248061653212177\\
5	0.17075573497431\\
10	0.146509691084213\\
15	0.14014800359335\\
20	0.137752413057114\\
25	0.137190510302906\\
30	0.1368782199127\\
};
\addlegendentry{$N_{RF}=15$}

\addplot [color=mycolor5, line width=2.0pt, mark=o, mark options={solid, mycolor5, mark size=3pt}]
  table[row sep=crcr]{%
0	0.17279389583239\\
5	0.0927743929702841\\
10	0.0575201493611395\\
15	0.0435218538078591\\
20	0.0399626314516202\\
25	0.0394452751208397\\
30	0.0387535956958196\\
};
\addlegendentry{$N_{RF}=20$}

\end{axis}
\end{tikzpicture}%
}
      \vspace{-3mm}
	 \caption{Channel estimation performances vs. uplink transmit power.} \label{ch_est_fig}
		\end{center}
  \vspace{-9mm}
\end{figure}
\vspace{-2mm}

\subsection{Channel Estimation Uncertainty}
We now assess the system's channel estimation performance, assuming that each UE has the same power budget $\mu$ and transmits at its maximum power.
The chosen performance metric is the normalized MSE averaged over all the subcarriers, which for UE $k$ is defined as $\textrm{MSE}_k=\sum_{v=1}^V\norm{\bb_{k,v}-\widehat{\bb}_{k,v}}^2/\norm{\bb_{k,v}}^2$.
Fig.~\ref{ch_est_fig} shows that the number of receiving RF chains plays a big role in the channel estimation performances, as bringing $\Nrf$ from $5$ to $10$ reduces the maximum channel estimation MSE of approximately $50\%$.
Furthermore, we can also notice a diminishing return when the number of receiving RF chains increases past $15$. This can be ascribed to the fact that with each additional RF chain, we add progressively weaker eigen-directions.
\vspace{-2mm}
\subsection{Positioning Performance}
\begin{figure}[t!]
\begin{center}
   \resizebox{0.44\textwidth}{!}{
    % This file was created by matlab2tikz.
%
%The latest updates can be retrieved from
%  http://www.mathworks.com/matlabcentral/fileexchange/22022-matlab2tikz-matlab2tikz
%where you can also make suggestions and rate matlab2tikz.
%
\definecolor{mycolor1}{rgb}{0.00000,0.44700,0.74100}%
\definecolor{mycolor2}{rgb}{0.85000,0.32500,0.09800}%
\definecolor{mycolor3}{rgb}{0.92900,0.69400,0.12500}%
\definecolor{mycolor4}{rgb}{0.49400,0.18400,0.55600}%
\definecolor{mycolor5}{rgb}{0.46600,0.67400,0.18800}%
\definecolor{mycolor6}{rgb}{0.30100,0.74500,0.93300}%
\begin{tikzpicture}[spy using outlines={rectangle, magnification=5,connect spies}]

\begin{axis}[%
width=0.5\textwidth,
height=2.193in,
at={(0.758in,0.499in)},
scale only axis,
xmin=-40,
xmax=0,
xlabel style={font=\color{white!15!black}},
xlabel={$\delta^2$[dB]},
ymode=log,
ymin=0.01,
ymax=11.3950062279738,
ylabel shift=-3pt,
legend columns=1,
yminorticks=true,
ylabel style={font=\color{white!15!black}},
ylabel={worst-case PEB [m]},
axis background/.style={fill=white},
xmajorgrids,
ymajorgrids,
yminorgrids,
legend style={nodes={scale=0.9, transform shape},at={(1.0,1.0)},legend cell align=left, align=left, draw=white!15!black}
]
\addplot [color=mycolor1, line width=1.5pt, mark=triangle, mark options={solid, mycolor1,mark size=3pt}]
  table[row sep=crcr]{%
-40	0.971307617001087\\
-35	0.605241144840322\\
-30	0.330130260503367\\
-25	0.223430359391616\\
-20	0.126565753947437\\
-15	0.0767709789257679\\
-10	0.0451249284538117\\
-5	0.0246076061939895\\
0	0.0137583141797587\\
};
\addlegendentry{$\TTheta$ optimized}

\addplot [color=mycolor4, line width=1.5pt, mark=o, mark options={solid, mycolor4,mark size=3pt}]
  table[row sep=crcr]{%
-40	10.4986975551732\\
-35	5.9043040120896\\
-30	3.320314723239\\
-25	1.86716443132718\\
-20	1.04998626118479\\
-15	0.590451116209861\\
-10	0.332035143705467\\
-5	0.186717097089433\\
0	0.104998742421002\\
};
\addlegendentry{$\TTheta$ random}

\addplot [color=mycolor3, line width=1.5pt, mark=square, mark options={solid, mycolor3,mark size=3pt}]
  table[row sep=crcr]{%
-40	11.3950062279738\\
-35	6.40823811202404\\
-30	3.60368029735061\\
-25	2.02650958764456\\
-20	1.13959208616757\\
-15	0.640840078836266\\
-10	0.360370922284972\\
-5	0.202651473181335\\
0	0.113959299965923\\
};
\addlegendentry{no RIS, $\Mrx=4$}

\addplot [color=mycolor2, line width=1.5pt, mark=diamond, mark options={solid, mycolor2,mark size=3pt}]
  table[row sep=crcr]{%
-40	3.1595260613787\\
-35	1.77673951956946\\
-30	0.999135379451211\\
-25	0.561855348843415\\
-20	0.315954523260648\\
-15	0.177674292768196\\
-10	0.0999135985939241\\
-5	0.0561855458710343\\
0	0.0315954544199824\\
};
\addlegendentry{no RIS, $\Mrx=25$}

\coordinate (spypoint) at (axis cs:-20,1.1);
\coordinate (spyviewer) at (axis cs:-30
,0.05);
\spy[width=3cm,height=2cm] on (spypoint) in node [fill=white] at (spyviewer);

\end{axis}
\end{tikzpicture}}
      \vspace{-3mm}
	 \caption{Worst-case PEB vs. RCS power for different RIS configurations. } \label{PEB_2}
		\end{center}
  \vspace{-9mm}
\end{figure}
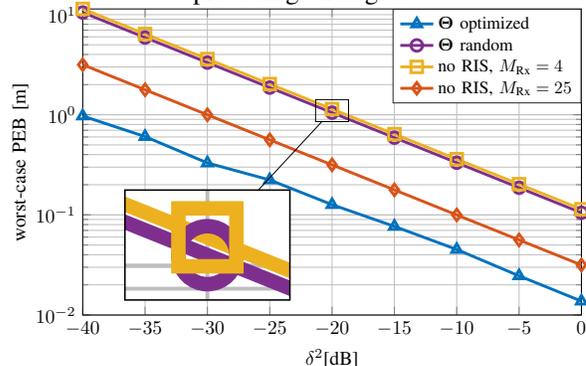

We now assess the positioning performance of the system by comparing it against an ideal scenario, which is labeled ``perfect'' in Fig.~\ref{PEB_1}. 
Here, the system minimises the PEB solely at one point, corresponding to the target location: this serves as a benchmark to assess the impact of imperfect prior knowledge on the system's localization accuracy.
Fig.~\ref{PEB_1} shows the performance degradation introduced by the system's limited prior knowledge of the target position, and that the inclusion of derivative beams greatly improves the positioning accuracy.
It is interesting to notice how $\WW^\textrm{S}$ and $\WW^\textrm{SE}$ produce similar performance, hinting at the fact that adding more directional beams does not produce a performance gain.
On the other hand, Fig.~\ref{PEB_2} shows how an unoptimized RIS produces the same performance as a RIS-less array. Interestingly, using only $\Mrx = 4$ active antennas with an RIS yields better performance than employing $\Mrx = 25$ active antennas without RIS integration. This highlights the strong potential of the optimized RIS-integrated array in minimizing the PEB.
\vspace{-6mm}
\section{conclusion}
\vspace{-2mm}
We have investigated the positioning performance of a bistatic JRC network where an RIS-integrated array is deployed as a sensing receiver.
In conclusion, numerical results demonstrate that incorporating derivative beams into the precoding matrix significantly enhances positioning accuracy, while an unoptimized RIS offers no performance improvement over conventional arrays. In contrast, optimized RIS-assisted configurations can achieve superior performance even with a substantially smaller number of active antennas.

\vspace{-2mm}

\begin{figure}[t!]
\begin{center}
   \resizebox{0.44\textwidth}{!}{
    % This file was created by matlab2tikz.
%
%The latest updates can be retrieved from
%  http://www.mathworks.com/matlabcentral/fileexchange/22022-matlab2tikz-matlab2tikz
%where you can also make suggestions and rate matlab2tikz.
%
\definecolor{mycolor1}{rgb}{0.00000,0.44700,0.74100}%
\definecolor{mycolor2}{rgb}{0.85000,0.32500,0.09800}%
\definecolor{mycolor3}{rgb}{0.92900,0.69400,0.12500}%
\definecolor{mycolor4}{rgb}{0.49400,0.18400,0.55600}%
\definecolor{mycolor5}{rgb}{0.46600,0.67400,0.18800}%
\definecolor{mycolor6}{rgb}{0.30100,0.74500,0.93300}%
\begin{tikzpicture}[spy using outlines={rectangle, magnification=10,connect spies}]

\begin{axis}[%
width=0.5\textwidth,
height=2.193in,
at={(0.758in,0.499in)},
scale only axis,
xmin=-40,
xmax=0,
xlabel style={font=\color{white!15!black}},
xlabel={$\delta^2$[dB]},
ymode=log,
ymin=0.005711382190564047,
ymax=1.60771003424303,
ylabel shift=-3pt,
legend columns=2,
yminorticks=true,
ylabel style={font=\color{white!15!black}},
ylabel={worst-case PEB [m]},
axis background/.style={fill=white},
xmajorgrids,
ymajorgrids,
yminorgrids,
legend style={nodes={scale=0.9, transform shape},at={(1.0,1.0)},legend cell align=left, align=left, draw=white!15!black}
]
\addplot [color=mycolor1, line width=1.5pt, mark=triangle, mark options={solid, mycolor1,mark size=3pt}]
  table[row sep=crcr]{%
-40	0.971307617001087\\
-35	0.605241144840322\\
-30	0.330130260503367\\
-25	0.223430359391616\\
-20	0.126565753947437\\
-15	0.0767709789257679\\
-10	0.0451249284538117\\
-5	0.0246076061939895\\
0	0.0137583141797587\\
};
\addlegendentry{$\WW^\textrm{SD}$}

\addplot [color=mycolor1, line width=1.5pt, mark=triangle, dashed, mark options={solid, mycolor1,mark size=3pt}]
  table[row sep=crcr]{%
-40	0.42114374945656\\
-35	0.236826548522344\\
-30	0.133177357725598\\
-25	0.0748911322852012\\
-20	0.0436686038141897\\
-15	0.0246603722872925\\
-10	0.0168720419180281\\
-5	0.00826273995522087\\
0	0.00587350627608914\\
};
\addlegendentry{$\WW^\textrm{SD}$ perfect}

\addplot [color=mycolor2, line width=1.5pt, mark=diamond, mark options={solid, mycolor2,mark size=3pt}]
  table[row sep=crcr]{%
-40	1.59942846887535\\
-35	0.907682154524086\\
-30	0.561565374969686\\
-25	0.39076176513445\\
-20	0.219132078363286\\
-15	0.1224335442222\\
-10	0.0661672717476757\\
-5	0.0390569922472064\\
0	0.0204835196786032\\
};
\addlegendentry{$\WW^\textrm{S}$}

\addplot [color=mycolor2, line width=1.5pt,dashed, mark=diamond, mark options={solid, mycolor2,mark size=3pt}]
  table[row sep=crcr]{%
-40	0.805485438724925\\
-35	0.452957862957\\
-30	0.254716945403524\\
-25	0.148918175334703\\
-20	0.0831027725676054\\
-15	0.0483785462002066\\
-10	0.0269409715035407\\
-5	0.0162187780653887\\
0	0.00951220057973956\\
};
\addlegendentry{$\WW^\textrm{S}$ perfect}

\addplot [color=mycolor5, line width=1.5pt, mark=square, mark options={solid, mycolor5,mark size=3pt}]
  table[row sep=crcr]{%
-40	1.60764826000189\\
-35	0.91566537902842\\
-30	0.57899077058591\\
-25	0.381620652572311\\
-20	0.218885982640876\\
-15	0.128617884608806\\
-10	0.0855798955180838\\
-5	0.0392233116798576\\
0	0.025120330012166\\
};
\addlegendentry{$\WW^\textrm{SE}$}

\addplot [color=mycolor5, line width=1.5pt, mark=square,dashed, mark options={solid, mycolor5,mark size=3pt}]
  table[row sep=crcr]{%
-40	0.796936257042485\\
-35	0.448150301012212\\
-30	0.25201345366229\\
-25	0.145247169503661\\
-20	0.0864423889118097\\
-15	0.0514081437102546\\
-10	0.0302951981318673\\
-5	0.0177515321835894\\
0	0.0104557800682374\\
};
\addlegendentry{$\WW^\textrm{SE}$ perfect}

\end{axis}
%----------------------------------------------------------------------------------------------------
%----------------------------------------------------------------------------------------------------%----------------------------------------------------------------------------------------------------
%----------------------------------------------------------------------------------------------------
% \begin{axis}[%
% nodes={scale=0.5, transform shape},
% width=0.15\textwidth,
% height=0.93in,
% at={(3.1858in,1.7499in)},
% scale only axis,
% xmin=-10,
% xmax=-5,
% xlabel style={font=\color{white!15!black}},
% xlabel={},
% ymin=1.23,
% ylabel shift=-10pt,
% ymode=log,
% legend columns=1,
% ymax=2.49099306240767,
% yminorticks=true,
% ylabel style={font=\color{white!15!black},font size=\tiny},
% ylabel={},
% axis background/.style={fill=white},
% % xmajorgrids,
% % ymajorgrids,
% % yminorgrids,
% legend style={nodes={scale=0.8, transform shape},at={(0.14,0.52)},legend cell align=left, align=left, draw=white!15!black}
% ]

% \addplot [color=mycolor2, line width=1.5pt, mark=o, mark options={solid, mycolor2,mark size=3pt}]
%   table[row sep=crcr]{%
% -10	2.26544417414325\\
% -5	1.27406359309997\\
% };

%\end{axis}
% \node (start) at (7.1,4.2) {};
% \node (end) at (8.2,4.6) {};
% \path[->] (start) edge [very thick] (end);
\end{tikzpicture}%
}
      \vspace{-3mm}
	 \caption{Worst-case PEB vs. RCS power for different precoding strategies. } \label{PEB_1}
		\end{center}
  \vspace{-8mm}
\end{figure}
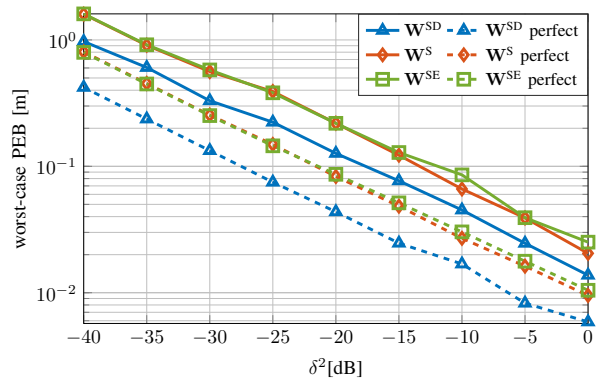

 \bibliographystyle{IEEEtran}
\bibliography{IEEEabrv, auxiliary/biblio}
\end{document}